\documentclass{stavanger}

\usepackage[utf8]{inputenc}
\usepackage{amssymb,amsmath,amsfonts,amsthm}
\usepackage{ dsfont }
\usepackage{hyperref}
\usepackage{natbib}
\usepackage{graphicx}
\usepackage{tikz}
\usepackage{cleveref}
\usepackage{mathtools}
\usepackage{placeins}
\usepackage{bm}
\usepackage{animate}
\usepackage[export]{adjustbox} 

\startlocaldefs

\endlocaldefs

\begin{document}


\begin{frontmatter}

\begin{fmbox}


\dochead{Preprint}{FP}


\title{A non-hermitean momentum operator for the particle in a box}

\author[
   addressref={aff1},                   	  
   email={skim@sejong.ac.kr}   		  
]{\inits{SK}\fnm{Seyong} \snm{Kim}}
\author[
   addressref={aff2},                   	  
   corref={aff2},                     		  
   email={alexander.rothkopf@uis.no}   		  
]{\inits{AR}\fnm{Alexander} \snm{Rothkopf}}


\address[id=aff1]{
  \orgname{Department of Physics}, 	 
  \street{Sejong University},                     		 
  \postcode{143-747},                               			 
  \city{Seoul},                              				 
  \cny{South Korea}                                   				 
}

\address[id=aff2]{
  \orgname{Faculty of Science and Technology}, 	 
  \street{University of Stavanger},                     		 
  \postcode{4021},                               			 
  \city{Stavanger},                              				 
  \cny{Norway}                                   				 
}



\end{fmbox}


\begin{abstractbox}
\begin{abstract} 
We construct a discrete non-hermitean momentum operator, which implements faithfully the non self-adjoint nature of momentum for a particle in a box. Its eigenfunctions are strictly limited to the interior of the box in the continuum limit, with the quarter wave as first non-trivial eigenstate. We show how to construct the corresponding hermitean Hamiltonian for the infinite well as concrete example to realize unitary dynamics. The resulting Hilbert space can be decomposed into a physical and unphysical subspace, which are mutually orthogonal. The physical subspace in the continuum limit reproduces that of the continuum theory and we give numerical evidence that the correct probability distributions for momentum and energy are recovered.
\end{abstract}



\end{abstractbox}


\end{frontmatter}


\section{Introduction}

\subsection{Physics motivation}

Our understanding of point particle motion rests on the concepts of position and momentum. In classical mechanics they constitute independent observables and future motion can be deduced uniquely and deterministically from their values at present. In quantum mechanics both quantities retain their role as the fundamental building blocks to describe motion of a point particle. In the microscopic domain their values however cannot be known simultaneously (at least not in the same spatial direction) and predictions about future motion become probabilistic in nature.

Remarkable phenomena occur when microscopic motion is confined between impenetrable walls. An explicit example are quantum dots, where one can observe the formation of bound states that are classically forbidden \cite{Schult:1989zz}. Another phenomenon is chaotic wave dynamics in quantum billiard, one example \cite{Seba:1990zz} being a triangular domain with a central point scatterer, which, in classical mechanics avoids chaotic behavior.

It is well known (see e.g. discussion in \cite{al2021alternative}) that the standard concept of momentum in quantum mechanics is unable to consistently describe motion in a geometry of finite extent. To be concrete, let us take a look at the one-dimensional infinite well. The conventional momentum operator $\hat p$ has the plane waves as eigenfunctions. If we measure momentum on a stationary state contained within the infinite well, i.e. we carry out a von Neumann projective measurement, then we expect to end up in one of the eigenstates of the momentum operator. These however are extended over the whole spatial domain and thus are incompatible with the finite extent of the well. This fact has been interpreted as the momentum measurement imparting an infinite amount of energy to the state, which is unphysical.

Interestingly the measurement of momentum most commonly is implemented via the measurement of position after the point particle is allowed to interact with its environment. Two examples are observing its position after free fall in a gravitational field or after being forced on a curved trajectory by a magnetic field. In each case, even though the particle may be confined in a finite geometry at first, it is released for the purpose of the momentum measurement (for a more comprehensive discussion see \cite{freericks2023measure}). 

Given the central role that momentum plays in our understanding of motion and the rich phenomenology of quantum systems in finite geometries, it is imperative that a consistent definition of this observable is achieved. In a recent study \cite{al2021alternative} a novel momentum operator was proposed. The authors construct an operator, with eigenfunctions, which are located exactly within the finite domain, solving a central shortcoming of the conventional definition of quantum mechanical momentum. On the other hand this momentum operator contains terms of infinite value located on the boundary in the form of delta-function contributions. In addition the Hamilton operator deployed in that study does not arise naturally from a combination of the momentum operator and the position operator.

In this study we set out to construct an alternative momentum operator $\hat p$ for the particle in the one-dimensional finite box in order to address these issues. Its eigenfunctions are also limited to the interior of the accessible domain. As it is agnostic of the boundary conditions of the stationary states, the operator is devoid of delta-function like contributions on the boundary. We show how it can be used to construct a consistent Hamiltonian, involving only reference to position $\hat x$ and momentum $\hat p$. The price we have to pay is that our momentum operator is not hermitean and does not lend itself to the conventional interpretation of a von Neumann projective measurement. We will give numerical evidence that our proposed momentum operator provides the correct statistical predictions of particle motion, as it reproduces the n-point functions of momentum $\langle \hat p^n\rangle$ with respect to the stationary states of the infinite well. 

\subsection{Observables in Quantum Mechanics}

In quantum mechanics observables are conventionally described by normal operators $\hat N\in {\cal N}$ with the defining property $\hat N \hat N^\dagger = \hat N^\dagger \hat N$. This important class of operators is not necessarily self-adjoint but is characterized by a purely real spectrum, i.e. the eigenvalues $\lambda_i$ in $\hat N |\lambda_i\rangle = \lambda_i |\lambda_i\rangle$ fulfill $\lambda_i \in {\mathbb R}$. This property allows normal operators to represent measurable properties of quantum systems. 

To interpret the measurement process as a von Neumann projective measurement, one reexpresses the normal operator representing the observable into a set of projection operators constructed from its eigenstates $|\lambda_i\rangle$
\begin{align}
    \hat N=\sum_i \lambda_i | \lambda_i\rangle \langle \lambda_i| = \sum_i \lambda_i \hat P_i,
\end{align}
where $\hat P_i \hat P_j = \hat P_i \delta_{ij}$. This decomposition is always possible as long as $\hat N$ is normal.

When carrying out the measurement of $N$ on a quantum system represented by the state $|\psi\rangle$, experimental evidence tells us that the state of the system spontaneously changes into one of the eigenstates of $\hat N$: $|\psi\rangle\to |\lambda_i\rangle$ and the measurement apparatus returns the value $\lambda_i$. The probability for this \textit{collapse of the wavefunction} into the state $|\lambda_i\rangle$ is given by $|c_i|^2= |\langle \lambda_i| \psi \rangle|^2$ if the observable can be represented by a normal operator $\hat N$.

A single measurement however does not allow us to confirm the probability $|c_i|^2$, and instead we must repeat the measurement a large number of times on identically prepared states and record a histogram of outcomes. Only after we have established knowledge of the probability distribution as a whole, is the information of an individual $|c_i|^2$ relevant. (i.e. each individual measurement will result in a purely random outcome that cannot be predicted. We can only aposteriori confirm that $\lambda_i$ has occurred with a frequency among our measurements in agreement with $|c_i|^2$.) 

Another way to think about the probabilistic nature of measurements in quantum mechanics is to consider not the probability distribution of outcomes $P(\lambda)$ itself, as we did before, but its moments. The lowest moments of a distribution are well known quantities, its mean $\langle \lambda \rangle = \int d\lambda \lambda P(\lambda) $ and its spread $\langle \lambda^2 \rangle = \int d\lambda \lambda^2 P(\lambda) $. These moments are straight forwardly related to the observable through its so called n-point functions $\langle \psi | \hat N^n | \psi \rangle$, where $n$ denotes the power of the operator $\hat N^n$ inserted in the inner product of the state $|\psi\rangle$. As discussed in the context of the Hamburger moment problem \cite{shohat1950problem}, under well-defined circumstances knowledge of all n-point functions is equivalent with the knowledge of the probability distribution. One prerequisite is that the probability distribution exists, which quantum mechanics guarantees.

In this paper we will study an operator on a discrete grid as candidate for a momentum operator for the particle in the box. This operator will not be normal. While it implements various properties of the continuum momentum operator faithfully, it does not allow a similar straight forward interpretation of measurement as a von Neumann projective measurement.

On the other hand we will find that it is possible to use our novel momentum operator to define from it a hermitean Hamilton operator for the infinite square well. This Hamiltonian encodes as part of its  eigenfunction space the physical energy eigenstates in the continuum limit. Consistency is found when inspecting the lowest n-point functions of the momentum operator evaluated in the physical energy eigenstates. Their values reproduce those of the continuum theory in the continuum limit. In turn we argue that our non-hermitean and non-normal momentum operator is a viable candidate to represent the physical property of momentum for a particle in a box.

\section{Momentum operator for a particle in a box}

\subsection{Defining properties}

When a microscopic point particle is restricted to a finite geometry of length $L$ with physical boundaries, its motion is affected by those boundaries. In order to describe a point particle in quantum mechanics we must define the position $\hat x$ and momentum $\hat p$ operator of the system at hand, from which the Hamiltonian operator $\hat H$ may be constructed. It is the Hamiltonian, which through functional exponentiation $\hat U = {\cal T} {\rm exp}[-i/\hbar\int dt \hat H]$ generates the dynamical unitary evolution that describes the motion of the point particle.

Central properties of operators in quantum mechanics are established through their behavior in the inner product of the system Hilbert space. When investigating these properties in a finite geometry, we choose to work in the position basis, where the scalar product is defined via an integral over wavefunctions $\langle x|\psi \rangle=\psi(x)$ and the position space matrix elements of the observable $\langle x|\hat N|y\rangle = N$ 
\begin{align}
 \langle \psi | \hat N  \chi \rangle = \int \, dx dy \; \psi^* (y) N (x,y) \chi(x)\label{eq:innerprodx}
\end{align}
When being placed in a finite geometry, the position operator $\hat x$ is not affected. By definition, it remains diagonal in the position basis, it is hermitean, self adjoint and its eigenfunctions remain localized in space.

On the other hand the momentum operator $\hat p$ differs from that of a system without boundaries. The reason lies in the fact that even though the momentum operator matrix elements in position space $p(x,y) = -i\hbar \delta(x-y) \frac{d}{dx}$ are local, the presence of the derivative means that their behavior within the inner product is affected by the presence of genuine boundaries
\begin{align}
    \langle \hat p \psi |  \chi \rangle &= \int_a^b \, dx \; \big( -i\hbar \frac{d}{dx} \psi(x)\big)^*  \chi(x)\\
    &= \int_a^b \, dx \; \psi(x)^* \big(-i\hbar \frac{d}{dx} \chi(x)\big) + \left. i \hbar \psi(x)^*\chi(x)\right|_a^b\\
    &= \langle  \psi |  \hat p \chi \rangle + \left. i \hbar \psi(x)^*\chi(x)\right|_a^b = \langle \hat{p}^\dagger \psi | \chi \rangle + \left. i \hbar \psi(x)^*\chi(x) \right|_a^b \label{eq:nonhermpgeomdef}
\end{align}
In other words, the momentum operator only remains hermitean, as long as one imposes periodic boundary conditions or restricts the domain of wavefunctions to those that vanish on the boundary $\psi(x)^*\chi(x)|_a^b=0$\footnote{Note that the momentum operator is invariant under spatial translations and thus does not carry explicit knowledge of the absolute coordinates of the geometry and thus it cannot distinguish between a box extending between $[-L/2,L/2]$ and $[0,L]$.}. 

Naively transplanting the continuum momentum operator into a finite geometry presents a conceptual challenge. Its eigenfunctions are plane waves, which extend over the whole of the real axis. In the conventional interpretation of projective measurement, this entails that a momentum measurement will collapse the wavefunction of a point particle within the geometry onto a wavefunction that is extended beyond the boundaries. In the explicit example of an infinite potential well, this would entail an infinite amount of energy being imparted on the particle. Such a picture hence must be unphysical.

\subsection{Summation by parts finite differences}

We thus wish to construct a momentum operator, which faithfully implements the behavior of the conventional momentum operator w.r.t. the scalar product of \cref{eq:nonhermpgeomdef}, but which possesses eigenfunctions which are limited to the accessible domain. 

Note that at this point we do not fix to any specific boundary condition, as this information must not enter the construction of the momentum operator. Understood as linear differential operator associated with a boundary value problem, the momentum operator and boundary conditions constitute two separate sets of information.

\begin{figure}
    \centering
    \includegraphics[scale=0.7]{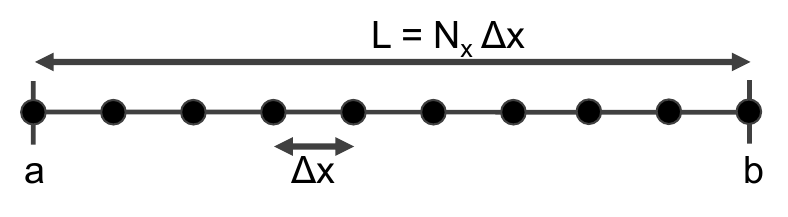}
    \caption{Sketch of the grid layout with genuine boundaries on which we formulate a momentum operator}
    \label{fig:gridlayout}
\end{figure}

Our strategy is as follows: to avoid technical intricacies of infinite dimensional vector spaces, we will begin our construction of a momentum operator on a discrete grid with $N_x$ points spaced $\Delta x=(b-a)/(N_x-1)$ apart and thus of total finite extend $L=(N_x-1)\Delta x$ as shown in \cref{fig:gridlayout}. 

The guiding principle for our construction is to require that the discrete operator mimics exactly the non-hermitian behavior of the continuum momentum operator manifest in \cref{eq:nonhermpgeomdef}. In addition we require that the operator becomes the generator of translations in the continuum limit, and thus must converge to the standard derivative as $N_x\to\infty$ and $\Delta x\to 0$ at constant $L$.

We take inspiration from modern discretization strategies for partial differential equations and classical actions of dynamical systems, for which the concept of a summation-by-parts (SBP) finite difference operator has been developed over the past decades (for reviews see e.g. \cite{svard2014review,fernandez2014review,lundquist2014sbp}). Summation-by-parts operators are designed to exactly mimic integration by parts in the discrete setting, one of our central requirements. 

Let us consider functions discretized on a grid ${\bm \psi}=(\psi(a),\psi(a+\Delta x),\psi(a+2\Delta x),\ldots,\psi(b))^{\rm T}$, where we choose without loss of generality $a=0$ and $b=1$. In order to mimic integration by parts we need a consistent definition of both integration and differentiation in the discrete setting.

As first step we thus select a specific quadrature rule for the integral representing the inner product in \cref{eq:innerprodx} via a diagonal and positive definite matrix $\mathds{H}$. The discrete inner product thus reads $({\bm \psi},{\bm \psi}^\prime)={\bm \psi}^{\rm \dagger} \mathds{H} {\bm \psi}^\prime$, where the symbol $\dagger$ refers to the combined transpose and complex conjugation.

A finite difference operator $\mathds{D}$ of order $r$ that mimics integration-by-parts (IBP) exactly, must obey the following properties: 
\begin{align}
     &\mathds{D} {\bm x}^n = n{\bm x}^{n-1} \quad {\rm for\, n<r}\label{eq:derivprop}\\
     &\mathds{D}=\mathds{H}^{-1}\mathds{Q}, \qquad \mathds{Q}^{\rm T}+\mathds{Q}=\mathds{E}_N-\mathds{E}_1={\rm diag}[-1,0,\ldots,0,1]\label{eq:derivsbp}.
\end{align}
Here \cref{eq:derivprop} establishes that $\mathds{D}$ indeed acts as an exact derivative on polynomials, as long as their degree is lower than $r$. \Cref{eq:derivsbp} on the other hand defines the structure of $\mathds{D}$ that, in the interplay with $\mathds{H}$, allows it to mimic IBP exactly in the discrete setting. The SBP operator is written as a product of the inverse quadrature matrix $\mathds{H}^{-1}$ and the matrix $\mathds{Q}$. The latter implements a central symmetric finite difference stencil in the interior, and stencils close to the boundary that allow it to fulfill the sumamtion-by-parts property given on the left of \cref{eq:derivsbp}. Summation-by-parts requires that when $\mathds{D}$ is moved within the inner product of the discretized function space, appropriate boundary terms arise. Such boundary terms make reference only to first and last entry of the discretized function vectors participating in the inner product\footnote{Note that this summation-by-parts property does not depend on the specific value of the discretized functions on the boundary or their functional form.}. Here projection operators $\mathds{E}_1$ and $\mathds{E}_N$ are used to express the summation-by-parts conditions. The former refers to a matrix with all entries zero except the top left, where it contain the value minus one. The latter is a matrix with only a single non-vanishing entry of value one in the bottom right corner. To be concrete, in the construction of our momentum operator, we deploy the lowest order diagonal norm SBP scheme, referred to as \texttt{SBP21}, which is second order in the interior and reduces to first order on the boundary. With the trapezoidal rule as choice for integral quadrature, we have
\begin{equation}
 \mathds{H}=\Delta x \left[ \begin{array}{ccccc} 1/2 & & & & \\ &1 & & &\\ & &\ddots && \\ &&&1&\\ &&&&1/2 \end{array} \right],
\quad 
\mathds{D}=
\frac{1}{2 \Delta x}
\left[ \begin{array}{ccccc} -2 &2 & & &\\ -1& 0& 1& &\\ & &\ddots && \\ &&-1&0&1\\ &&&-2&2 \end{array} \right].\label{eq:SBP21}
\end{equation}
Note that this construction differs from the one of Ref.~\cite{al2021alternative} and that our derivative operator does not need to be amended by terms specific to the boundary condition.

Let us also confirm that the finite difference operator $\mathds{D}$ is antisymmetric under the parity operation ${\cal P} \mathds{D}\to-\mathds{D}$ for ${\cal P} x\to-x$. This property follows from \cref{eq:SBP21}, as the only term that is affected by the change in sign of $x$ is the lattice spacing $\Delta x \to - \Delta x$.

\subsection{Spectral structure}

In preparation for using $\mathds{D}$ as building block to our momentum operator, let us investigate its spectrum. We follow here the strategy laid out in Ref.~\cite{ruggiu2020eigenvalue}. The presence of a different stencil in the interior and on the boundary in $\mathds{D}$ precludes us from using the simple ansatz of a single trigonometric function as eigenfunction. Instead, to determine the eigenvectors and eigenfunctions, one first solves the eigenvector relation for the interior stencil and exploits the necessary consistency with the boundary behavior as additional constraint for obtaining the discrete eigenvalues.

Without loss of generality, let $\Delta x=1$ for the remainder of this subsection. In \cite{ruggiu2020eigenvalue} the authors considered an SBP operator, which was amended by a so called simultaneous-approximation term (SAT). For this type of SBP-SAT operators there exists a theorem that guarantees that their spectrum is located on the imaginary axis. Our operator $\mathds{D}$ does not fulfill the same property. At the same time, we find from a heuristic numerical survey of the spectrum of $\mathds{D}$ for various choices of $N_x$ that its eigenvalues too are purely imaginary. For even values of $N_x$ one finds two zero eigenvalue, while for odd values of $N_x$ there exist three. This however does not represent a problem, as the associated eigenspace remains one-dimensional and, as expected, is spanned solely by the constant function.

We thus start our analysis using the same ansatz ${\bm \psi}_k=r^k$ as in \cite{ruggiu2020eigenvalue} to solve the internal stencil relation for the k-th entry of an eigenvector $\bm \psi$ with a purely imaginary eigenvalue $i\xi$ with $\xi\in\mathbb{R}$
\begin{align}
   \frac{1}{2} {\bm \psi}_{k+1}- \frac{1}{2} {\bm \psi}_{k-1}= i\xi{\bm \psi}_{k} \qquad \Rightarrow \qquad r^{k+1}-r^{k-1}=2i\xi r^k. \label{eq:intstenc}
\end{align}
We see that there are two possible solutions for the values of $r_{1,2}=i\xi\pm \sqrt{1-\xi^2}$, which imply $r_1r_2=-1$.

Assuming that $r_1\neq r_2$, i.e. $\xi\neq 1$ we are led to the following ansatz for the eigenvector as linear combination of two independent contributions
\begin{align}
    {\bm \psi}_{k}(\xi)=c_1(\xi)r_1^k(\xi)+c_2(\xi)r^k_2(\xi). \label{eq:solDevecans}
\end{align}
By inserting \cref{eq:solDevecans} into the boundary stencil relation at the left boundary involving ${\bm \psi}_{0}$ and ${\bm \psi}_{1}$ one finds that $c_1$ and $c_2$ can be expressed solely in terms of ${\bm \psi}_{0}$ and $\xi$ by proxy of $r_1$ and $r_2$
\begin{align}
   c^{\rm L}_1/{\bm \psi}_{0}=  -\frac{2 \xi ^2+i \xi  (r_2-2)+r_2-1}{r_1 (r_1-r_2)}, \quad  c^{\rm L}_2/{\bm \psi}_{0}= \frac{2 \xi ^2+i \xi  (r_1-2)+r_1-1}{r_2 (r_1-r_2)}.\label{eq:ccoeffL}
\end{align}
Writing $c_1$ and $c_2$ in the above fashion we exclude the case of a zero eigenvalue for which $r_1=r_2=0$. I.e. we will not recover $\xi=0$ from this analysis. This missing information is obtained from the construction of $\mathds{D}$, which tells us that the eigenvector with zero eigenvalue is simply the constant function.

Similarly we can insert the ansatz \cref{eq:solDevecans} into the right boundary relation to obtain
\begin{align}
   &c^{\rm R}_1/{\bm \psi}_{N_x}=  \frac{2r_2 \xi^2 +i \xi (2r_2 - 1) - r_2+1}{r_1^{Nx-2} (r_1-r_2)}, \nonumber \\
   &c^{\rm R}_2/{\bm \psi}_{N_x}=\frac{2 r_1 \xi^2 + i \xi (2r_1 - 1) -r_1+1}{r_2^{Nx-2} (r_2-r_1)}.
\end{align}

In order for us to obtain a consistent solution to the eigenvector relation both in the interior and on the boundary the solutions for ${\bm c}^{\rm L,R}= ( c^{\rm L,R}_1,c^{\rm L,R}_2)^T$ must represent the same vector, i.e. ${\bm c}^{\rm L}$ and ${\bm c}^{\rm R}$ must be linearly dependent. This requirement is most efficiently expressed in that the determinant of the $2\times 2$ matrix constructed of ${\bm c}^{\rm L}$ and ${\bm c}^{\rm R}$ must vanish
\begin{align}
    \left| \begin{array}{cc} c^{\rm L}_1/{\bm \psi}_{0} & c^{\rm L}_2/{\bm \psi}_{0}\\ c^{\rm R}_1/{\bm \psi}_{N_x} & c^{\rm R}_2/{\bm \psi}_{N_x} \end{array}\right| = 0.\label{eq:bndident}
\end{align}
Using the \texttt{Mathematica} software\footnote{Accompanying computations for this paper can be found at \cite{rothkopf_2024_10843823}.}, we solve the above equation and find by comparison to \cite{ruggiu2020eigenvalue} that for even values of $N_x$ the allowed $N_x-2$ eigenvalues $\xi\neq0$ for $\mathds{D}$ are given by
\begin{align}
    \xi_l={\rm sin}\Big[\big(\frac{l}{N_x-1}+\frac{1}{2}\big)\pi\Big]={\rm cos}\Big[\big(\frac{l}{N_x-1}\big)\pi\Big], \qquad l \in[1,\ldots,N_x-2].\label{eq:Deigenval}
\end{align}

Once the eigenvalues are known, we can construct the eigenfunctions by inserting the definition of the coefficients $c_{1,2}$ from e.g. \cref{eq:ccoeffL} into \cref{eq:solDevecans}. A thorough discussion of the structure of the eigenfunctions will be presented in the next section where $\mathds{D}$ is used to construct a candidate momentum operator.

\subsection{Momentum operator candidate and its properties}

Having introduced the summation-by-parts finite difference operator $\mathds{D}$ in the previous section, we proceed to construct from it a momentum operator $\hat p$. For finite grid spacing we will refer to its discrete form as $\mathds{P}$, from which the physical momentum operator $\hat p$ emerges in the continuum limit
\begin{align}
    \mathds{P} = -i\hbar \mathds{D}, \qquad{\rm where} \qquad \hat p = \lim_{\tiny\begin{array}{c}\Delta x\to0\\N_x\to\infty\\L=(N_x-1)\Delta x\end{array}} \mathds{P}.\label{eq:defp} 
\end{align}
Due to the fact that $\mathds{D}$ is odd under parity, also $\mathds{P}$ retains this property. Note that it is even under combined ${\cal PT}$ symmetry, as ${\cal T}$ sends $i\to=-i$.

Let us inspect its spectral properties. From the study of the spectrum of $\mathds{D}$, it follows via \cref{eq:defp} that the discrete eigenvalues $p_l$ of $\mathds{P}$ are purely real and  on a grid with even $N_x$ take on the following values
\begin{align}
    p_l = \left\{\begin{array}{lc}
    \frac{1}{\Delta x} {\rm sin}\Big[\big(\frac{l}{N_x-1}+\frac{1}{2}\big)\pi\Big] = \frac{1}{\Delta x} {\rm sin}\Big[\big(\frac{l \Delta x}{L} +\frac{1}{2}\big)\pi\Big] & l\in[1,\ldots,N_x-2]\\
    0& l=N_x-1,N_x.\end{array}\right.\label{eq:peigenval}
\end{align}
Let us consider what \cref{eq:peigenval} tells us about the spectrum of $\mathds{P}$. Due to the antisymmetry of the ${\rm sin}$ function around $\pi$, all the eigenvalues $p_l$ come in pairs of equal magnitude and opposite sign. The smallest non-vanishing magnitude of the momentum is obtained for $l=N_x/2$ and $l=N_x/2-1$. We find
\begin{align}
\nonumber p_{N_x/2}=&\frac{1}{\Delta x} {\rm sin}\Big[\big(\frac{N_x \Delta x}{2L} +\frac{1}{2}\big)\pi\Big]=\frac{1}{\Delta x} {\rm sin}\Big[\big(\frac{(N_x-1) \Delta x}{2L} + \frac{\Delta x}{2L} +\frac{1}{2}\big)\pi\Big]\\
=&\frac{1}{\Delta x} {\rm sin}\Big[\big( \frac{\Delta x}{2L} +1\big)\pi\Big] \overset{\Delta x/L\ll1}{\approx}- \frac{\pi}{2L} = -\frac{2\pi}{4L}.\label{eq:qrtwave}
\end{align}
Interestingly it turns out that the smallest momentum eigenvalue corresponds to a wavelength of $\lambda=4L$, which indicates that the associated eigenvector resembles a quarter wave, in contrast to the half-wave that has been obtained in previous studies, such as in \cite{al2021alternative}. As one lets the grid spacing in \cref{eq:peigenval} tend towards zero the largest accessible momentum value increases, until in the continuum limit the whole real axis of momenta is covered.

Note that the doubly degenerate zero mode is associated with a one-dimensional eigenspace, spanned by the constant function and thus does not pose a conceptual problem. We have confirmed by numerical determination of the spectrum, that for an odd number of grid points a third degenerate zero eigenvalue ensues. The space spanned by the associated eigenfunctions however still remains one-dimensional, as each zero eigenvalue is associated with the same function.

The appearance of multiple zero modes may be understood intuitively as follows. The spectra of $\mathds{P}$ and $\mathds{P}^{\rm T}$ are necessarily the same, as they are based on the same characteristic polynomial. However, since $\mathds{P}$ is non hermitean, its right and left eigenvectors do not necessarily agree. In particular it is known from the study of the underlying SBP finite difference operator in \cite{Rothkopf:2022zfb} that the right zero-mode is the constant function and the left zero-mode is the highly oscillatory $\pi$-mode. In order to accommodate such two different zero modes in the common spectrum, it needs to feature at least two zero eigenvalues. 

In case that we have an odd number of grid points, this $N_x$ determines the overall number of eigenvalues of $\mathds{P}$. Since all eigenvalues have to come in positive-negative pairs, in addition to the two zero modes, one is left with one remaining eigenvalue to fill. The only number that is its own negative is zero and thus one naturally expects to find another zero mode in case that $N_x$ is odd.

We can now visually inspect the eigenfunctions $\bm \psi$ of the operator $\mathds{P}{\bm \psi}_l=p_l {\bm \psi}_l$, which follow from inserting the definition of the coefficients $c_{1,2}$ from e.g. \cref{eq:ccoeffL} into \cref{eq:solDevecans} and using the values just obtained in \cref{eq:peigenval} as the eigenvalues.

\begin{figure}
    \centering
    \includegraphics[scale=0.3]{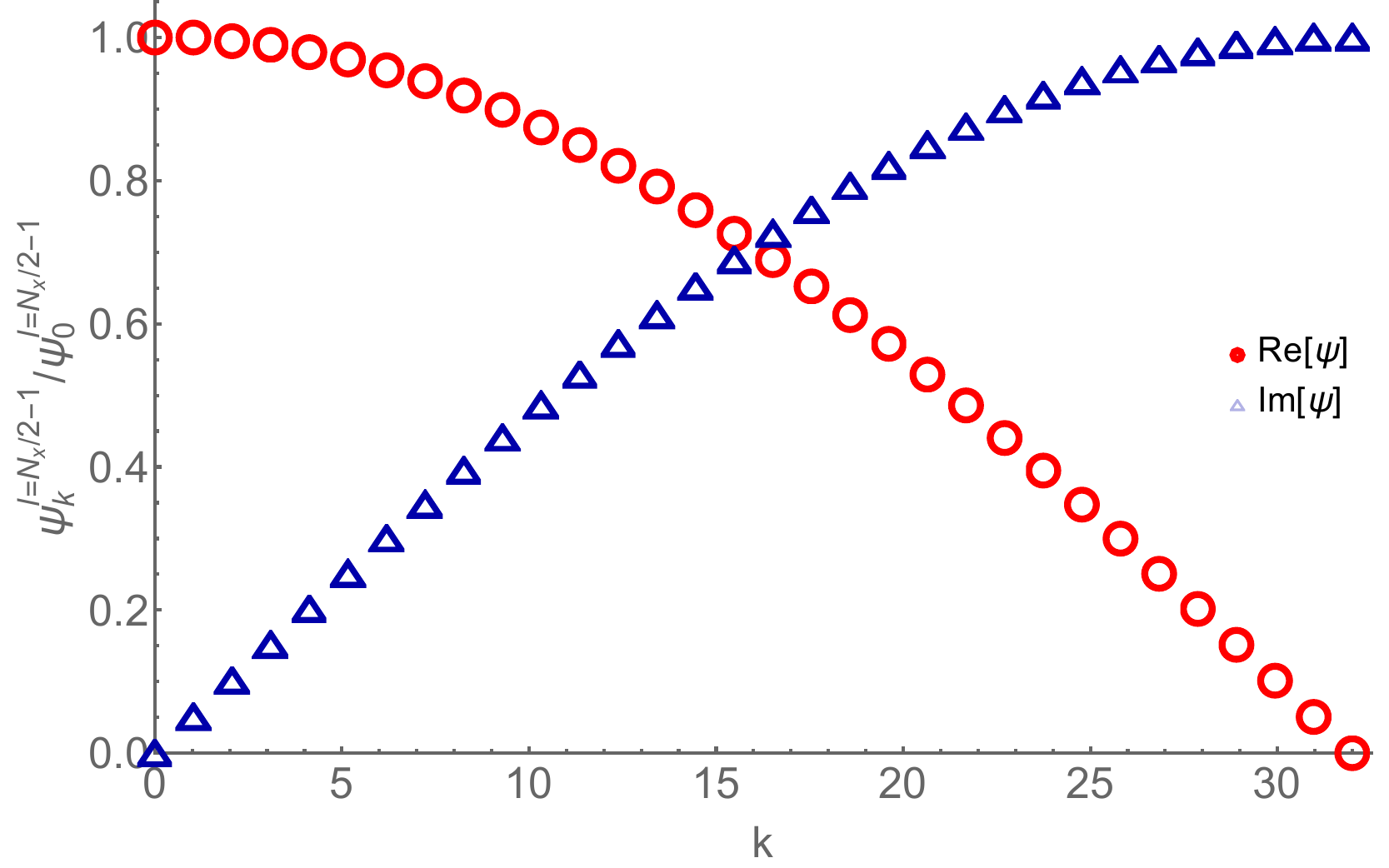}
    \includegraphics[scale=0.3]{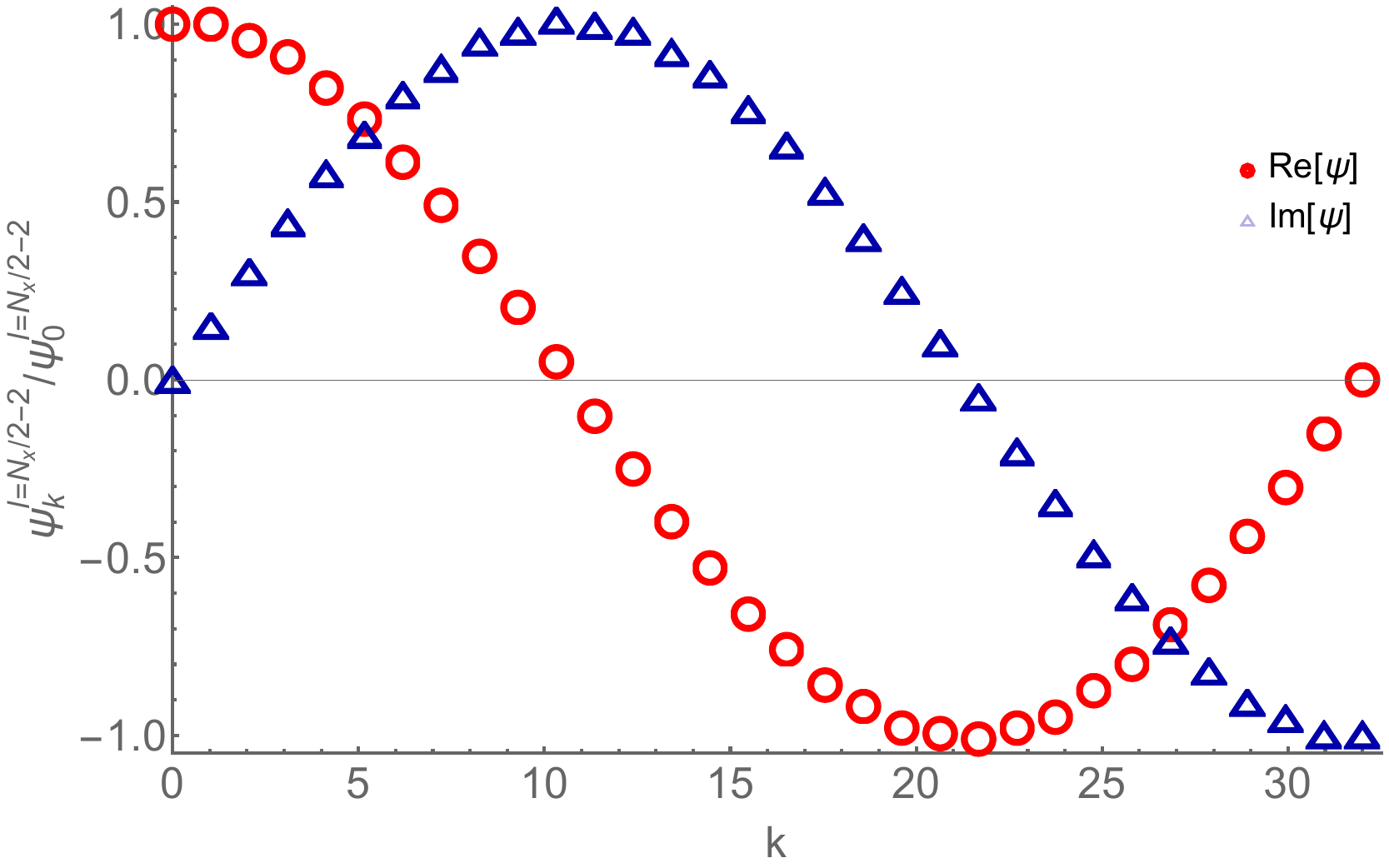}
    \caption{Example of two non-constant eigenvectors of $\mathds{P}$ in position space for $N_x=32$. Real-part (blue circles) and imaginary-part (mustard triangles) of the normalized eigenfunction to $l=N_x/2-1$ (top) and to $l=N_x/2-2$ (bottom). Note that on the boundary either the value, or the derivative of the functions is zero.}
    \label{fig:pEigenvec}
\end{figure}

In \cref{fig:pEigenvec} we plot the real- (red circles) and imaginary part (blue triangles) of the lowest non-trivial normalized eigenfunction, corresponding to the eigenvalue $l=N_x/2-1$ (top) and for the second to lowest non-trivial eigenfunction $l=N_x/2-2$ (bottom). 

While the boundary behavior entails that these functions are not exactly plane waves, they exhibit very similar oscillatory features in the interior of the domain. As predicted by \cref{eq:qrtwave}, we find that the lowest non-trivial eigenfunction of the operator $\mathds{P}$ corresponds to the quarter wave. Note that this structure is the smallest fraction of a standing wave that can be consistently placed within a bounded domain.

The condition that needs to be fulfilled by such an eigenfunction is that no probability leaks from the finite domain. We will now show that in the continuum limit this property is indeed fulfilled. Note that the real- and imaginary part of the eigenfunctions in \cref{fig:pEigenvec} either takes on the value zero at the boundary or, crucially, exhibits a zero derivative. This is most clearly seen in the lower panel of \cref{fig:pEigenvec} where the first two red circles on the left and the last two blue triangles on the right take on exactly the same value leading to a vanishing of the derivative there.

We can check the probability flux through the boundary defined as
\begin{align}
    j(x)=&\frac{1}{2mi}\Big( \psi^*(x)\frac{\partial}{\partial x} \psi(x) - \frac{\partial}{\partial x} \psi^*(x)\psi(x)\Big)\\
    \approx& \frac{1}{2mi} \Big( {\bm \psi}^* \circ {\mathds{D}}{\bm \psi} - {\mathds{D}}{\bm \psi}^* \circ{\bm \psi}\Big), 
\end{align}
where in the second line the symbol $\circ$ refers to component-wise multiplication. Boundary leakage is related to a non-zero value of $j$ at the boundaries of the domain. Since the reference to $m$ is made for purely dimensional reasons, we will set $m=1$ at this point.

\begin{figure}
    \centering
    \includegraphics[scale=0.35]{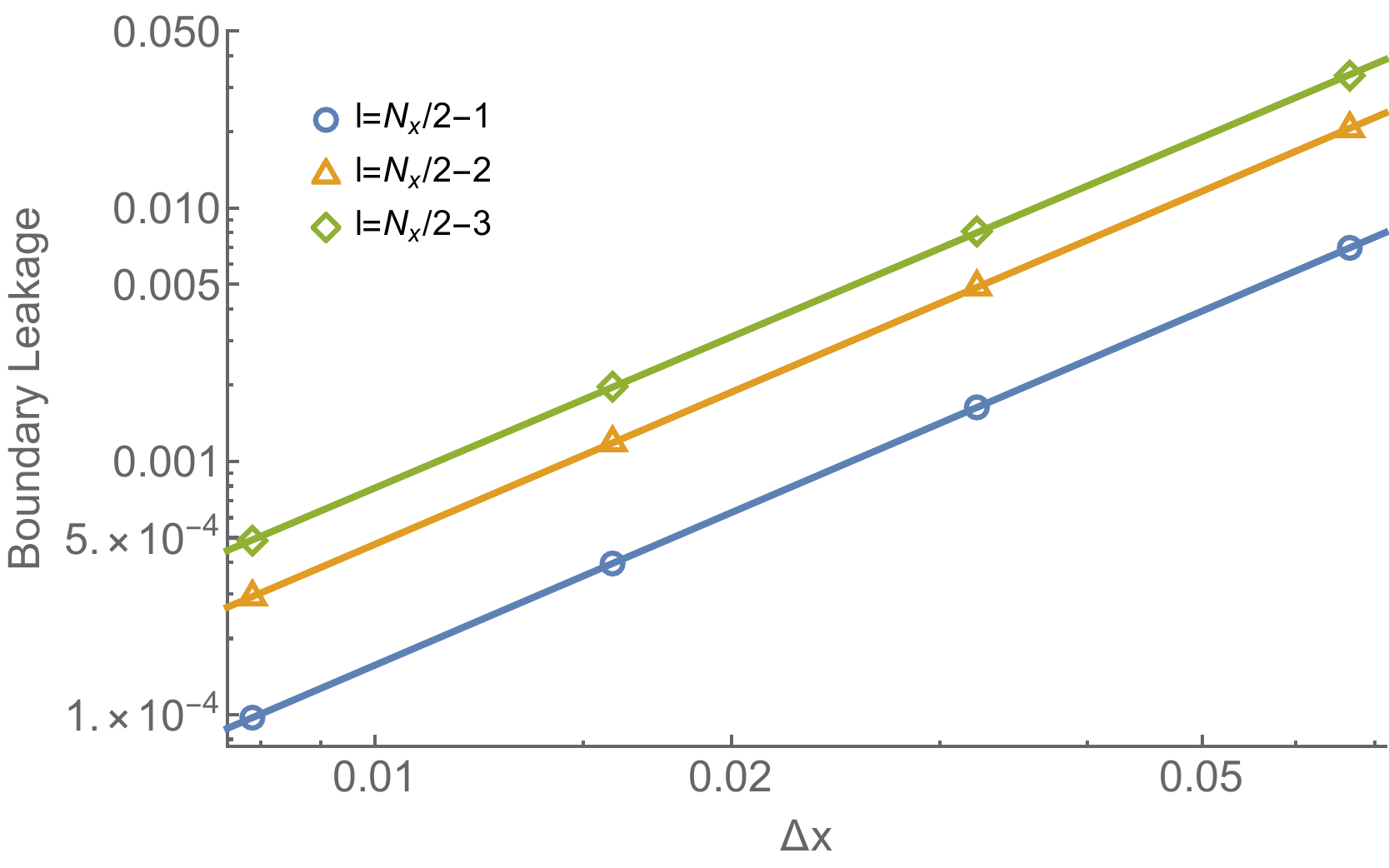}
    \caption{Disappearance of boundary leakage under grid refinement by the three lowest lying non-constant eigenfunctions of the operator $\mathds{P}$.}
    \label{fig:bndleakdis}
\end{figure}

In \cref{fig:bndleakdis} we show the remnant boundary leakage for the lowest three non-trivial eigenfunctions of $\mathds{P}$. One finds that even on relatively coarse grids with $N_x=16$, the right-most data points, the values we obtain a small. More importantly as one refines the grid, going to smaller values of $\Delta x$, the remnant boundary leakage diminishes with a powerlaw, according to $\Delta x^\nu$ with $\nu\approx 2$. It is reassuring that even with the lowest order \texttt{SBP21} finite difference operator a quadratic scaling of this crucial property towards the continuum limit is achieved.

By now we have confirmed two milestones in the construction of a viable candidate momentum operator for the particle in the box. Our discrete operator $\mathds{P}$ features eigenfunctions, which in the continuum limit are exactly constrained to the accessible domain. In addition the operator is defined without reference to any boundary conditions that may be enforced on the stationary states of the system and thus is devoid of the infinite constant boundary terms necessary in other constructions of alternative momentum operators (see e.g. \cite{al2021alternative}).

One may ask, how our operator will recover the well-known eigenfunctions of the momentum operator in an unbounded domain? If we naively let the size of the domain increase, by increasing the number of grid points, then, due to the normalization of the eigenfunctions, the quarter wave will be increasingly pulled apart and eventually become the zero function. The same would happen to all the other eigenfunctions with extent of a finite half-integer multiple of the wavelength. We know however that the (unnormalizable) eigenfunctions of the conventional momentum operator are the finite plane waves associated with a continuum of eigenvalues. 

This conundrum can be resolved by recognizing that there is a qualitative difference between the setup of finite extent with genuine boundaries and the infinite extend without explicit boundary terms. This difference formally manifests itself in \cref{eq:bndident}, where the presence of the explicit boundaries provides us with a quantization conditions for the  eigenvalues of the SBP finite difference operator and thus for our proposed momentum operator. In case that the domain is infinite we have no explicit boundary terms and thus \cref{eq:bndident} is not applicable. Indeed for finite lattice spacing $\Delta x$ but $N_x$ taken to infinity $\mathds{D}$ would only contain the interior symmetric stencil, considered in \cref{eq:intstenc}. The corresponding eigenvalue relation for the momentum operator, in the absence of boundary terms is solved by 
\begin{align}
\mathds{P}{\bm \psi}=p^{N_x=\infty} {\bm \psi}. \qquad {\bm \psi}_k = {\rm exp}[i \kappa \Delta x k], \qquad p^{N_x=\infty}=\frac{1}{\Delta x} {\rm sin}[\kappa \Delta x],
\end{align}
for continuous values of $-\frac{\pi}{\Delta x}<\kappa<\frac{\pi}{\Delta x}$.

These are exactly the unnormalizable plane waves that arise for the conventional momentum operator on the unbounded domain. Thus our novel momentum operator not only produces eigenfunctions of finite extent and without boundary leakage in the presence of genuine boundaries but also reproduces the known spectrum as the extend of the domain is let go to infinity.

Now we must confirm that the operator defined in this section can be used to consistently construct a quantum theory of point particle motion undergoing unitary time evolution, which is the focus of the next section.

\section{Hamiltonian for a particle in a box}

The defining properties of microscopic motion are encoded in the position $\hat x$ and momentum operator $\hat p$, which fulfill the canonical commutation relation
\begin{align}
    [\hat p,\hat x] = -i \hbar \hat I.
\end{align}
From these fundamental building blocks a hermitean or at least ${\cal PT}$ symmetric Hamiltonian is constructed, which as part of the time evolution operator generates unitary time evolution.

In our discrete setting the counterpart to $\hat x$ is $\mathds{X}={\rm diag}[a, a+\Delta x, \ldots,b]$. Together with our definition of $\mathds{P}$, we find that 
\begin{align}
    \Big(\mathds{P}\mathds{X}-\mathds{X}\mathds{P}\Big){\bm \psi}&= -i\hbar ( \psi_1, \frac{1}{2}(\psi_0+\psi_2),\frac{1}{2}(\psi_1+\psi_3),\ldots, \psi_{N_x-1})\\
    &= -i\hbar ( \psi_0 , \psi_1,\psi_2,\ldots, \psi_{N_x}) + {\cal{O}}(\Delta x).
\end{align}
That is, the commutation relation between $\hat x$ and $\hat p$ is recovered in the continuum limit. And while convergence overall is of linear order, one converges to the correct result quadratically in the interior.

The operator $\mathds{P}$ features a purely real spectrum, even though is is not described by a hermitean matrix $\mathds{P}^\dagger\neq \mathds{P}$. This requires care when using our momentum operator to define the corresponding point particle Hamiltonian. We suggest the following generalized construction
\begin{align}
    \mathds{H} = \frac{ \mathds{P}^\dagger \mathds{P}}{2m} + V(\mathds{X}), \qquad {\rm where}\qquad  \hat H = \frac{\hat p^\dagger \hat p}{2m} + V(\hat x) = \lim_{\Delta x\to0} \mathds{H},\label{eq:DefH}
\end{align}
which of course agrees with the conventional definition of the Hamiltonian in case that the momentum operator is hermitean. Note that we only make reference to the two operators $\mathds{X}$ and $\mathds{P}$ and do not need to introduce a separate Hamiltonian in addition to these (c.f. Ref.~\cite{al2021alternative}).

The non-standard Hamiltonian of \cref{eq:DefH}, by construction, is hermitean and thus leads to unitary time evolution. 

\subsection{Particle in the infinite well}

Let us focus on a model system, in which the finite extend of the spatial domain is vital: the infinite square well. This system is defined by the potential function
\begin{align}
    V(x)=\left\{ \begin{array}{lc} 0 & a<x<b \\ \infty & {\rm otherwise}\end{array}\right. \qquad {\rm approx.} \qquad V({\mathds{X}})={\rm diag}[10^7,0,\ldots,0,10^7],
\end{align}
which in the discrete setting is approximated with large but finite values on the boundary. In order to accommodate such a large dynamic range of values we use in the following the high-precision arithmetic of the \texttt{Mathematica} software, when investigating the spectral structure of $\mathds{H}$.

It is the presence of the potential function $V$, which enforces the finiteness of the domain in this case. In an analytic computation, if we require the wavefunction of our system to remain continuous, the presence of an impenetrable potential wall forces us to adopt boundary conditions where the wavefunction vanishes at the boundary. This continuous behavior is implied when solving for the eigenvectors based on the Hamilton operator $\mathds{H}$ defined above. I.e. as the value of the approximate potential function $V$ are increased the resulting eigenvectors approach the value zero on the boundary more and more closely.

\begin{figure}
    \centering
    \includegraphics[scale=0.3]{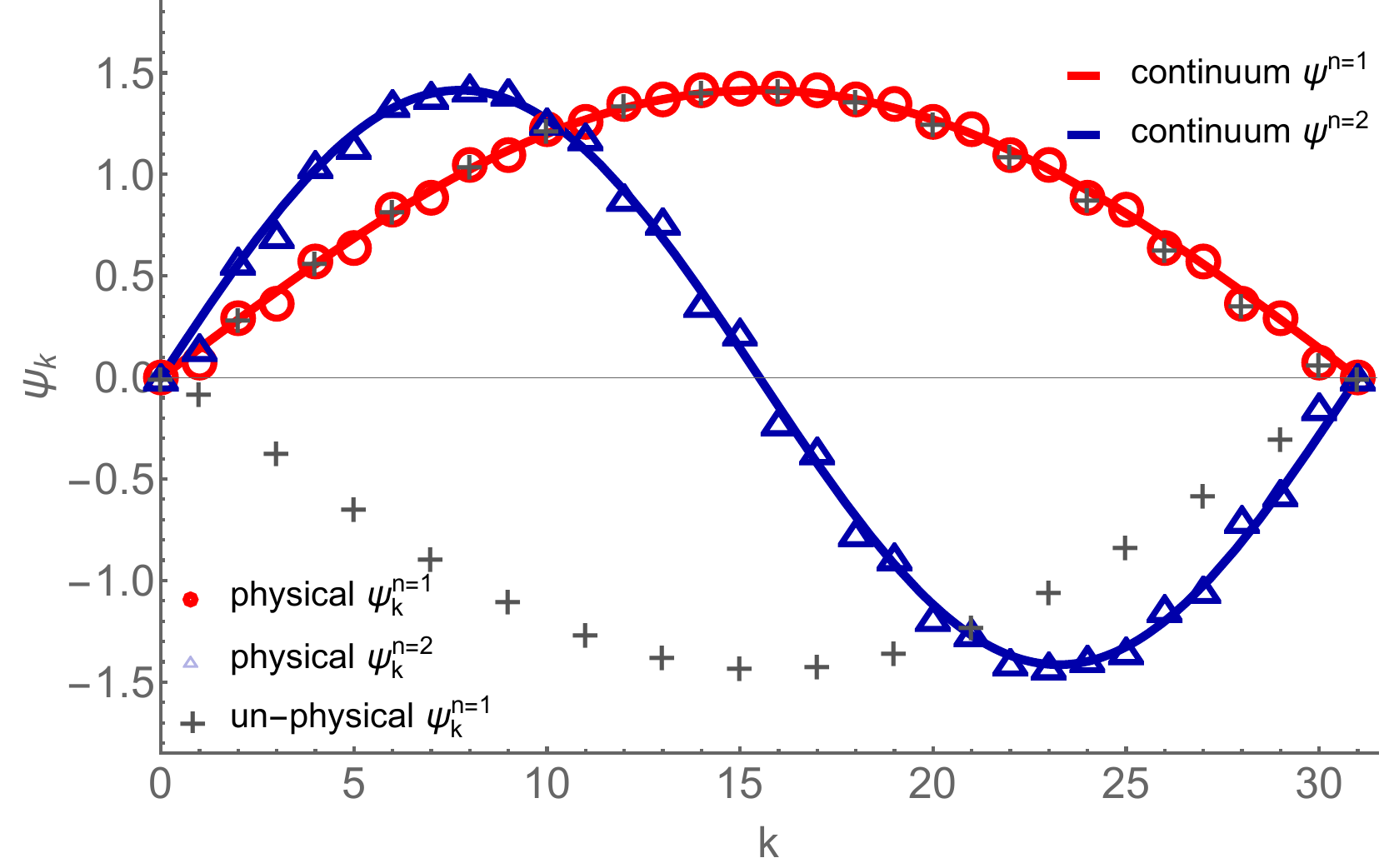}
    \caption{Examples of the two lowest lying physical (colored symbols) and unphysical (gray cross) energy eigenfunctions of the Hamilton operator $\mathds{H}$ built from our non-hermitean momentum operator $\mathds{P}$ on a $N_x=32$ grid. The continuum ground state (red) and first excited (blue) stationary state wavefunctions are provided as solid lines.}
    \label{fig:nrgeigen}
\end{figure}

When inspecting the spectrum of $\mathds{H}$, we find several peculiar properties. The first is that the spectrum $E_n$ is double degenerate, similar to the behavior observed in \cite{al2021alternative}. We can understand this phenomenon from recent insight into the spectral structure of summation-by-parts operators \cite{Rothkopf:2022zfb}. The fact that $\mathds{D}$ is not anti-symmetric means that its left and right eigenvectors are not necessarily the same. And indeed while the right eigenvector with zero eigenvalue is the physical constant zero mode, it turns out that the left eigenvector associated with a zero eigenvalue is the maximally oscillating function that can be resolved on a finite grid of $N_x$ points, the so called $\pi$-mode. The occurrence of such artificial oscillatory zero-modes is well known from the study of symmetric discretization schemes in lattice field theory and known there as the doubler-problem.

And since $\mathds{H}$ contains both $\mathds{P}$ and $\mathds{P}^\dagger$ its eigenfunctions $\mathds{H} | \psi^n\rangle = E_n |\psi^n\rangle$ are affected by the presence of such oscillatory modes. In \cref{fig:nrgeigen} we plot two of the lowest lying physical stationary states for $N_x=32$. The ground state as red circles and the first excited state as blue triangles. The continuum solutions for this system are provided as colored solid lines. We find that the physical stationary states are already well reproduced in the discrete setting.

In addition to the physical modes, we also plot as gray crosses one of the unphysical eigenfunctions, the one with the same energy eigenvalue as the physical ground state. This eigenfunction is exactly the point-wise product of the physical ground state wavefunction and the $\pi$-mode.

While in non-linear quantum field theory doubler-modes represent a formidable challenge, in linear quantum mechanics they are less malign. Indeed, since $\mathds{H}$ is hermitean, the physical and unphysical eigenstates are exactly orthogonal. In turn the Hilbert space of stationary states decomposes into a direct sum of a physical and unphysical Hilbert space. Importantly, unitary time evolution will not mix between the two subspaces, i.e. if the initial conditions of the time evolution are prepared from physical states, our $\mathds{H}$ will produce only physical states at later time.

\begin{figure}
    \centering
    \includegraphics[scale=0.3]{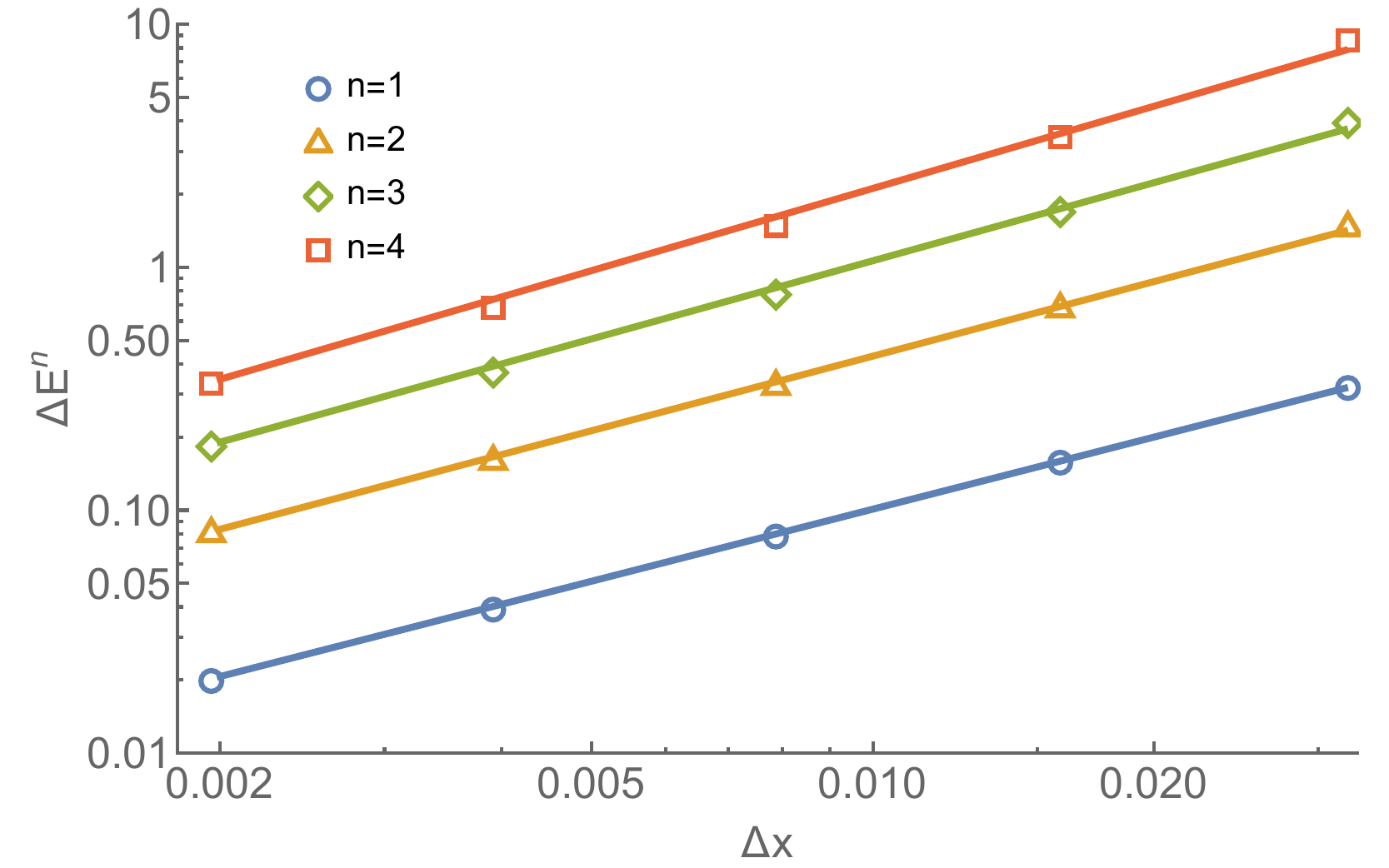}
    \caption{Deviation $\Delta E^n$ of the lowest four $n=1,\ldots,4$ energy eigenvalues of $\mathds{H}$ from the continuum values of the infinite square well for different lattice spacings $\Delta x$.}
    \label{fig:nrgeigenconv}
\end{figure}

Let us collect numerical evidence that the unconventional construction of $\mathds{H}$ reproduces the known results for the infinite well in the continuum limit. To this end we plot in \cref{fig:nrgeigenconv} the difference $\Delta E^n$ between the eigenvalue of the n-th stationary state obtained numerically from $\mathds{H}$ and the analytic continuum expression. As one sees, this difference diminishes consistently under refinement of the grid $\Delta x\to0$. The powerlaw dependence of $\Delta E^n\propto \Delta x^\nu$ with $\nu\approx1$ tells us that here the influence of the boundary remains relevant, since it is only there that the $\mathds{P}$ converges linearly to the continuum limit. (Note that since we use an approximate potential function, at some point deviations from the continuum result for the infinite potential well will appear as $\Delta x$ is made smaller than the values considered here.)

We find that our $\mathds{H}$, constructed from $\mathds{P}$, reproduces, within the physical subspace of its Hilbert space, the correct stationary states in the continuum limit. This further supports $\mathds{P}$ as candidate for an appropriate momentum operator for the particle in a finite box.

\subsection{Momentum and Energy in the finite domain}

We now turn to a discussion of momentum measurement in the infinite potential well, carried out on a initially prepared stationary state. 

Since our momentum operator is not normal, due to its structure at the boundary, its eigenfunctions are not mutually orthogonal nor complete. While one can always orthogonalize the set of eigenvectors, the fact that the sum over their outer product does not reproduce the identity operator means that $\mathds{P}$ does not lend itself to the conventional interpretation of a von Neumann projective measurement. I.e. a stationary state cannot be uniquely decomposed into the set of eigenstates of the momentum operator. Can we still make predictions about momentum measurements using $\mathds{P}$?

As alluded to in the introduction, we may either make predictions on the outcome of experiments by citing the probabilities of individual measurements, or we may refer to the set of n-point functions of the operator encoding the observable of interest. Each provide us with complementary insight on the probabilistic nature of measurement. 

Note however that only expectation values are deterministic predictions of quantum mechanics. I.e. what can in principle be predicted without uncertainty is the mean of outcomes of a large number of experiments carried out on an ensemble of identically prepared states, as the number of measurements is taken to infinity. Thus knowledge of the n-point functions of the momentum operator constitutes all possible deterministic predictions quantum mechanics can make about momentum of the point particle.

Interestingly the evaluation of $\langle \psi| \hat p|\psi \rangle$ only makes reference to the local properties of the wavefunction ${\psi}$, as the momentum operator in position space is a derivative. Since we are guaranteed that $\mathds{D}$ converges to the correct derivative $d/dx$ in the continuum limit, we expect that we are able to reproduce the continuum momentum expectations values by use of $\mathds{P}$. 

\begin{figure}
    \centering
    \includegraphics[scale=0.3]{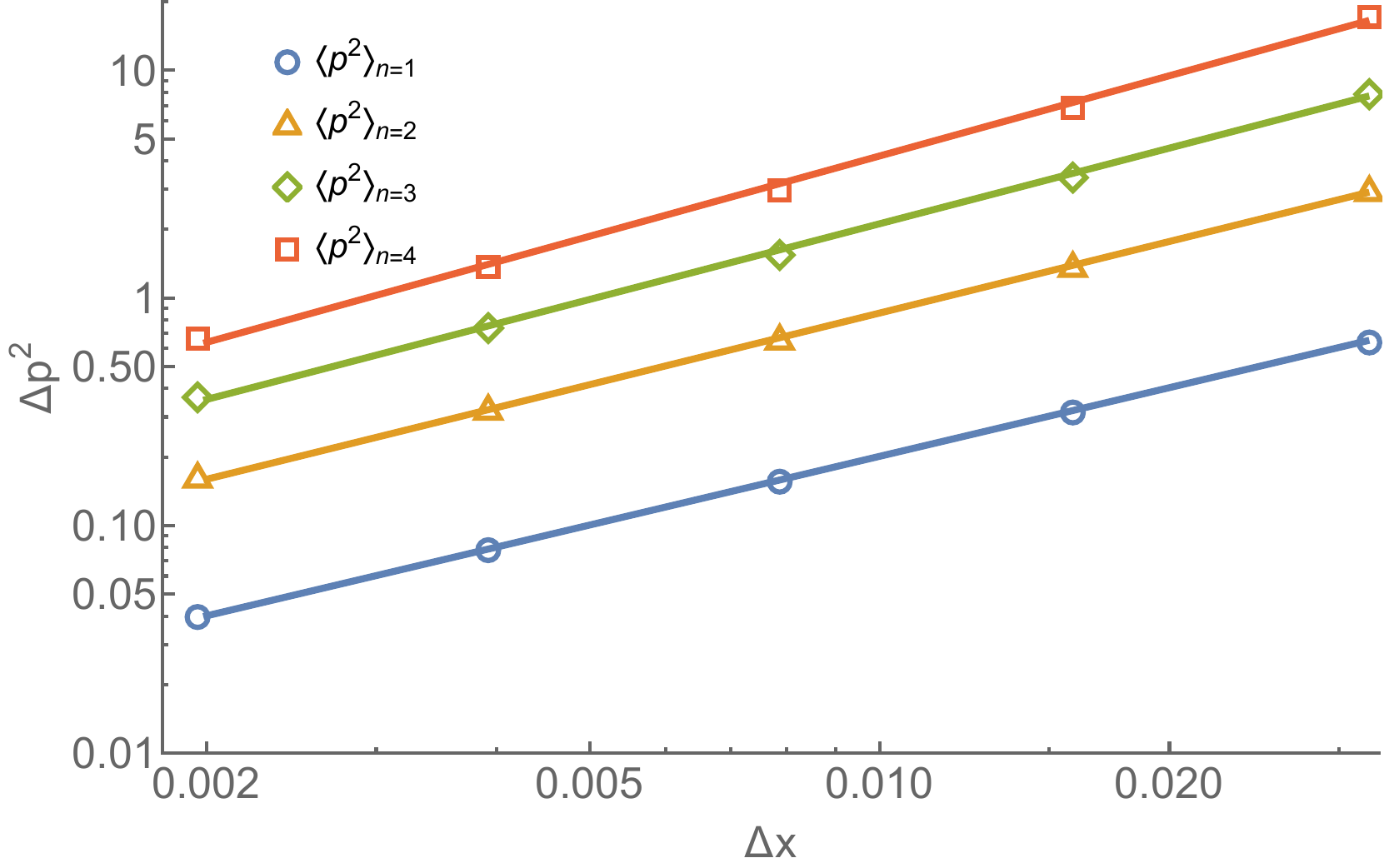}
    \caption{Diminishing of $\Delta p^2$ the deviation of the momentum two-point function from its continuum value evaluated in the lowest four energy eigenstates $n=1,\ldots,4$.}
    \label{fig:p2nrgeval}
\end{figure}

\begin{figure}
    \centering
    \includegraphics[scale=0.3]{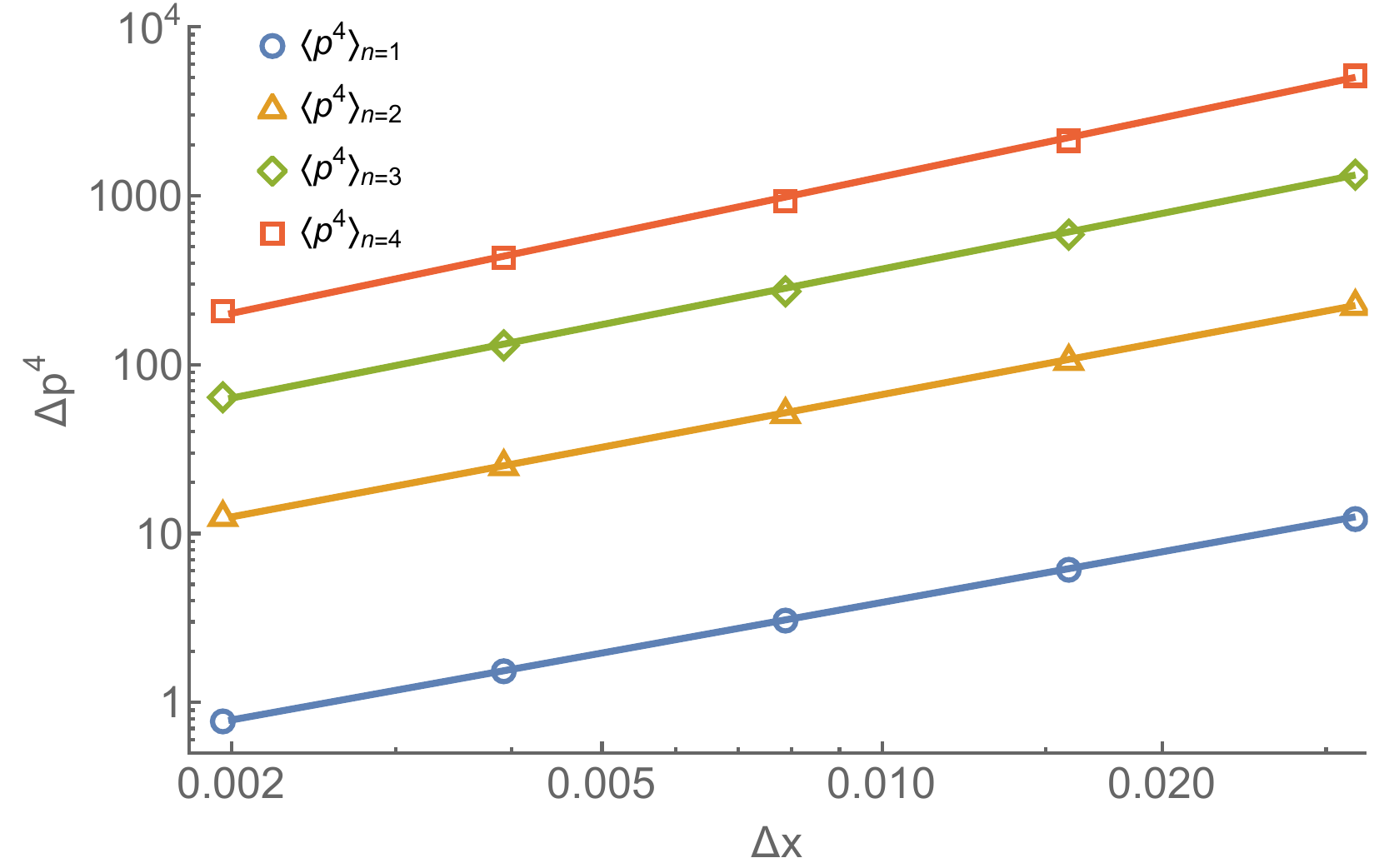}
    \caption{Diminishing of $\Delta p^4$ the deviation of the momentum four-point function from its continuum value evaluated in the lowest four energy eigenstates $n=1,\ldots,4$.}
    \label{fig:p4nrgeval}
\end{figure}

And indeed, as shown in \cref{fig:p2nrgeval} and \cref{fig:p4nrgeval}, we find numerical evidence that the correct n-point functions ensue. In \cref{fig:p2nrgeval} we plot the difference $\Delta p^2=\langle \mathds{P}^2 \rangle-\langle \hat p^2\rangle$ between the value of $\langle \mathds{P}^2 \rangle$ in the four lowest lying stationary states (ground state data as open circles) obtained from $\mathds{H}$ and the continuum expectation value $\langle \hat p^2 \rangle$ for the infinite well. The same comparison is made in \cref{fig:p4nrgeval} for the four-point function $\Delta p^4 = \langle \mathds{P}^4 \rangle-\langle \hat p^4 \rangle$.

While the absolute differences on coarse grids may at first sight appear relatively large, they consistently diminish towards zero as the grid is refined $\Delta x\to0$. We find for both $\langle \Delta p^2 \rangle$ and $\langle \Delta p^4 \rangle$ a powerlaw $\propto \Delta x^\nu$ with $\nu\approx 1$. The odd n-point function are, as required by symmetry, exactly zero.

The lowest four n-point functions contain a wealth of phenomenologically relevant information about the observable of interest and \cref{fig:p2nrgeval,fig:p4nrgeval} suggest that we are able to reproduce them reliably in our setup based on $\mathds{P}$ and $\mathds{H}$.

There is another unusual feature in the relation of the momentum operator and the Hamiltonian in a finite domain: they do not commute. In \cite{al2021alternative} this fact has been interpreted in the context of the von Neumann projective measurement. In that context a measurement of $\hat p$ will project the stationary state of the well into one of its eigenstates. These eigenstates however are not simultaneous eigenstates of the Hamiltonian and actually constitute linear combinations from both physical and unphysical parts of the Hilbert space. Since the unphysical states are associated with infinite energy solutions, the momentum measurement was interpreted to endow the point particle with an infinite amount of energy.

As we do not have a similar interpretation of the measurement as a von Neumann projective measurement in our setup, we would like to provide a different interpretation of the fact that $[\mathds{H},\mathds{P}]\neq0$. Making only reference to the operators themselves, we may inspect Heisenberg's equations of motion, which states
\begin{align}
    \frac{d}{dt} \mathds{P} = \frac{1}{i\hbar} [\mathds{H},\mathds{P}] \neq0.
\end{align}
In an unrestricted domain, where the system is spatially translationally invariant, momentum must be preserved, i.e. the time derivative of $\mathds{P}$ must vanish. This is the case for the conventional momentum operator and the Hamiltonian in the absence of a potential. In the presence of impenetrable walls, or more generally in the presence of a $\hat x$ dependent potential term, translational invariance is lost. An encounter with the walls of the well reverses the momentum of the particle. For an infinite well this finite change in momentum happens instantaneously. 

Thus in Heisenberg's equation of motion the commutator between $\mathds{P}$ and $\mathds{H}$ necessarily must lead to a nonzero value and since we are considering the infinite well it actually must lead to an infinite value due to the instantaneous nature of the reversal of momentum at the boundary. Thus we consider the non-commutativity of momentum and Hamiltonian as a necessary consequence of the absence of translation symmetry.

\section{Summary}

We have presented a concrete realization of a momentum operator for a point particle confined to a finite domain. Starting from a discretized setting, we use a summation-by-parts finite difference operator $\mathds{D}$ to construct a corresponding momentum operator $\mathds{P}$, which mimics exactly the boundary effects associated with integration by parts in the inner product of wavefunctions.  

We investigated the spectral properties of this momentum operator confirming that its eigenfunctions are indeed confined to the interior of the domain in the continuum limit, where no residual boundary leakage occurs. To connect to the known spectrum of the momentum operator on the infinite domain, we discussed how the absence of genuine boundaries leads to an eigenvalue relation that produces the correct plane wave solutions.

The fact that $\mathds{P}$ is described by a non-hermitean matrix, required us to resort to a non-standard construction of the Hamiltonian for the point particle involving $\mathds{P}$ and $\mathds{P}^\dagger$. We showed that the spectrum of the Hamiltonian contains the physical stationary states and we reproduce the correct energies of the infinite well in the continuum limit. 

The symmetric stencil used in the interior of the operator $\mathds{D}$ leads to the occurrence of oscillatory doubler-modes in the spectrum of the Hamiltonian. These however are orthogonal to the physical states and unitary time evolution does not mix between the two sectors.

Finally we confirmed that the four lowest n-point functions of the operator $\mathds{P}$ evaluated in the low-lying eigenstates of $\mathds{H}$ reproduce the continuum values as the lattice spacing is reduced $\Delta x\to0$. 

And while relevant statistical information about momentum measurement is contained within the correctly reproduced n-point functions we cannot use $\mathds{P}$ to make statements about individual projective measurements, as it is not normal, and thus it cannot be uniquely decomposed in projection operators made from its eigenvectors.

Last but not least, we relate the fact that Hamiltonian and momentum do not commute to the loss of translation symmetry and the encounter of the point particle with the impenetrable wall, which leads to an instantaneous reversal of its momentum.

\subsection{Further venues for investigation}

The spectrum of the momentum operator $\mathds{P}$ differs from previously suggested candidate operators in that the first non-trivial eigenfunction is the quarter wave. It would be interesting to see, whether different momentum concepts can be distinguished in experiment according to the lowest possible momentum value to be observed. It has to be kept in mind though that (at least) on the level of the lowest four expectation values, $\mathds{P}$ makes exactly the same predictions as the continuum $\hat p$.

One may wish to also revisit the question of how to consistently combine solutions of the time independent Schr\"odinger equation if they are considered to arise in separated domains connected by interfaces, similar to the study in \cite{al2021canonical}. The solution of the finite well, e.g. proceeds by dividing space up into three domains, in each of which one would have to deploy a momentum operator similar to $\mathds{P}$, adapted to the existence of a genuine boundary.

In addition it remains to be determined whether and how a path integral could be formulated based on the concept of a non-hermitean momentum operator. This question is directly related to the use of summation-by-parts finite difference operators for the discretization of the system action in the functional approach to quantum mechanics and quantum field theory.

We believe that our proposed candidate for a momentum operator on a finite domain $\mathds{P}$ together with a non-standard construction of an associated hermitean Hamiltonian $\mathds{H}$ provides a new viewpoint to this active field of research and hope that it will contribute to the eventual resolution of this fundamental question in quantum mechanics.

\section*{Acknowledgements}
A.~R. acknowledges fruitful discussions with David Ploog. This research project was initiated as part of the Korea-Nordic R\&D networking grant 2022K2A9A2A23000252. S.~K is supported by the National Research Foundation of Korea, under grant NRF-2021R1A2C1092701.


\appendix    

\FloatBarrier

\begin{backmatter}

\section*{Competing interests}
  The authors declare that they have no competing interests.

\section*{Author's contributions}
    \begin{itemize}
         \item S.~Kim: literature review, infinite volume limit, discussions, editing, grant management
         \item A.~Rothkopf: construction of the momentum operator, computation of spectrum, numerical implementation, writing, editing
    \end{itemize}


\bibliographystyle{stavanger-mathphys}


\bibliography{references}


\end{backmatter}


\end{document}